# WRITING ELECTRONIC DEVICES ON PAPER WITH CARBON NANOTUBE INK

M. Dragoman,[1*] E. Flahaut,[2] D. Dragoman,[3] M. Al Ahmad,[4] R. Plana[4]

[1]National Research and Development Institute in Microtechnology, Str. Erou Iancu Nicolae 32B, 077190 Bucharest, Romania

[2]Centre Interuniversitaire de Recherche et d'Ingénierie des Matériaux, UMR CNRS 5085, Université Paul Sabatier, 31062 Toulouse, France

[3]Univ. Bucharest, Physics Dept., P.O. Box MG-11, 077125 Bucharest, Romania

[4] LAAS CNRS, 7 Avenue du Colonel Roche, 31077 Toulouse Cedex 4, France

**The normal paper used in any printer is among the cheapest flexible organic materials that exist. We demonstrate that we can print on paper high-frequency circuits tunable with an applied dc voltage. This is possible with the help of an ink containing functionalized carbon nanotubes and water. After the water is evaporated from the paper, the nanotubes remain steadily imprinted on paper, showing a semiconducting behaviour and tunable electrical properties.**

The paper is the most encountered and cheapest flexible organic material. Moreover, paper is a dielectric up to 3 GHz, with an electric permittivity of $\varepsilon_r \cong 3.3$ and losses expressed by $\tan\delta \cong 0.08$. Using inkjet printing technology and silver nanoparticle inks, some basic RF circuits, such as antennas, resonators, and wireless identification tags (RFID) were implemented and tested [1]. On the other hand, inkjet printing on various substrates is extensively used to fabricate various patterns or devices based on carbon nanotube (CNT) inks. In this respect, multi-walled carbon nanotube (MWCNT) inks were used to print various patterns on plastic and paper. After tens of times repetitive printing with a common printer with the cartridge loaded with the MWCNT ink, in the dried ink the random network of MWCNTs attached on the substrate displayed an impedance with the real part of few MΩ up to 1 MHz. However, no impedance tuning with the applied voltage was observed [2]. CNT transistors based on various ink solutions were also printed on various organic substrates or semiconducting doped Si substrates used as back gate using inkjet techniques [3]. Moreover, single-walled carbon nanotubes (SWCNT) conductive lines up to 75 mm in length were fabricated using inkjet printing on glass substrates; their real part of the impedance decreases from 517 kΩ to 7.4 kΩ at 100 kHz as the number of overwriting increases from one to eight [4].

The aim of this paper is to merge the researches and technologies of paper as a high frequency substrate with the CNT ink technology to implement high-frequency devices. A prerequisite of almost any RF device is to be tunable, i.e. the real part ($R$) or the imaginary part ($X$) of the impedance $Z$, or both, must be tunable when a dc bias voltage ($V_b$) is applied. This is a feature never met up to now in RF devices on paper, although encountered in previous studies on random MWCNTs placed between nanogaps [5].

The characteristic $Z(V_b) = R(V_b) + iX(V_b)$ is crucial for the implementation of high-frequency devices on paper substrates, since the $R = (V_b)$ and $X(V_b)$ dependences are unavoidable features for matching of RF networks, for bandwidth tuning of filters, and RF signal detection. In principle, no radio or sensor in the RF spectrum is possible without having in its configuration tunable components such resistors, capacitances or inductances. When these dependences are nonlinear, RF detection or mixing is possible. So, the quest of a CNT ink deposited on paper, which could produce conductive lines or patterns with a tunable impedance opens the possibility to build tags, sensors, and even a radio on paper. This would be a significant advance of wireless technology based on nanotechnologies.

The search of the CNT ink printable on normal paper, with printed patterns that display a tunable impedance, started with the choice of the CNT type to be used in ink. We have known from previous studies that double-walled carbon nanotube (DWCNT) mixtures, i.e. randomly oriented DWCNTs and SWCNTs as in the case of the CNT inks, have a dielectric behaviour up to 65 GHz [6]. Up to 3 GHz the electric permittivity of DWCNTs is about 4 and $\tan\delta \cong 0.7$. The further step was then to functionalize the DWCNTs shown in Fig. 1 to display a semiconducting behaviour and to be hydrophilic in water.

The DWCNTs were prepared by CCVD synthesis as described earlier [7]. We obtained samples that contain typically ca. 80% of DWCNT, together with 15% of SWCNT and 5% triple-walled CNT [7]. A stable suspension of DWCNTs without the need of the addition of a surfactant was prepared by rendering the surface of the DWCNT hydrophilic via an oxidation treatment. More precisely, 50 mg of DWCNT were oxidized by heating for 3 h in 35 mL of 0.38M solution of $K_2Cr_2O_7$ in $H_2SO_4$ 8.2M. The setup was equipped with a condenser to avoid the evaporation of the liquid during the operation. After oxidation, the sample was first filtered on a polypropylene membrane (with a pore size of 0.45 mm) and then washed with deionised water until neutrality. Afterwards, the sample was dried in air at 80°C in an oven. To prepare a suspension in water, 3 mg of oxidized DWCNT were mixed together with 5 mL of deionised water and dispersed by sonication (bath) for 10 min. After a 12 h sedimentation, the supernatant was separated.

The TEM (JEOL 1011 operated at 100 kV) observation of the oxidized DWCNT, represented in Fig. 2, revealed that some of the CNTs have probably been cut. The presence of an amorphous-like deposit on the DWCNT is obvious and possibly corresponds to hydrolyzed oxygenated functions or graphene-like carbon debris.

The oxidized DWCNT were characterized by Raman spectroscopy (see Fig.3). Five spectra were recorded in different places of the sample with $\lambda = 632nm$. The mean ratio intensity between the D and G bands was close to 30%, which suggests the presence of structural defects in the CNT due to the oxidizing treatment. The presence of RBM peaks confirms that in spite of the strong oxidizing conditions, DWCNT are still present in the sample. Further, as illustrated in Fig.4, we have prepared different CNT inks denoted A to F in which the DWCNT were introduced in water in different concentrations, and an additional ink G in which the DWCNT were introduced in NMP (N-methyl-2-pyrrolidone). The ink A has the highest concentration of CNTs, of 1 mg/L, the CNT concentrations decreasing progressively up to F; the CNT concentration in the ink G is 10 mg/L. Then, with a pipette we have added the same number of CNT ink drops on normal paper, forming a disk with a diameter of 7 mm. We have measured the impedance of all CNT ink disks after the water was evaporated, and we repeated the measurements after several hours, and then days. After two months the disks formed by the CNT ink deposited on paper were not dispersed and the impedance measurements were only slightly different from those presented below.

The impedance was measured with the impedance analyzer HP 4294 A connected to a probe station, which has two sharp needles located at the margins of the CNT disks. The impedance analyzer was calibrated with a calibration test set before the measurements. The CNT disks are denoted with the same letters as the ink from which they have originated. Since the diameter of the disks are much longer than the length of the CNTs, the measured current is due to the formation of a random network of semiconducting CNTs. Measurements showed that CNT disks A to E have an impedance tunable with the applied voltage, while the disk G has displayed a very small resistance and no tunability. In all cases the CNT concentration is higher the threshold needed for the formation of a percolation path for conduction through carrier hopping between the semiconducting CNTs. In Fig. 5 we present the results on the CNT disk A, Figs. 5a and 5b displaying, respectively, the dependence of $R$ and $X$ on frequency at various applied voltages. These figures show that the resistance decreases as a function of the applied voltage, being almost constant at applied bias of 10 V (with a value of 30 k$\Omega$) as a function of frequency up to a few MHz, while the reactance has the behavior of a tunable capacitance, i.e. a varactor, beyond 1 MHz. Lower values of the resistances could be obtained by increasing the number of imprints, as indicated also in Ref 4. The voltage dependence of the resistance is a clear indication of the semiconductor nature of the random networks. The increase of the capacitance at high frequencies could be related to the mechanism reported in Ref. 2, which involves a capacitive coupling between CNTs at high frequencies. The instabilities observed in both the resistance and (especially) the reactance dependence on frequency above 10 MHz are consistent with a similar behavior reported in Ref. 4, due to insufficient network formation in the sample.

As the concentration of CNTs in the disks decrease, the resistance should increase. The dependence on the applied voltage of the resistance and the reactance of disk E at 1 MHz is represented in Fig. 6. The resistance is higher than that of disk A at higher voltages, and its dependence on the applied dc bias is less pronounced. From Fig.6 it follows also that the reactance of the disk E behaves as a tunable inductance. This behavior is consistent with that described in Ref. 2, where it is shown that the inductive behavior can be justified by regarding the curved nanotubes as tiny coils. This behavior is stronger for low CNT densities, since at higher CNT concentrations the local magnetic fields in the coils affect each other and the total inductance decreases.

How will look a RF radio on paper? Any direct receiver should have an antenna which could be a folded dipole or another antenna configuration, a tunable filter, and a nonlinear element for detection and  batteries.  In principle, except the battery which must be mounted separately on paper, the entire radio could be printed on paper using a normal printer since the antenna [1], the tunable filter and the detector can be printed on paper using the techniques described above. The radio on paper will consist of an antenna (which will be the subject of further researches, but which could be printed on paper using silver nanoparticles ), and a couple of disks containing CNT ink with various concentrations. The connections between various variable devices could be made with CNT ink of concentration G which has an impedance mostly resistive of around 600 Ω  which does not depend strongly on the applied voltage. This concept of radio on paper is illustrated in Fig. 7 where, in the case of inks A and E, playing the role of detector and filter, respectively, we have introduced their photos. In conclusion, the paper demonstrated that inexpensive tags, sensors or even a rudimentary radio could be fabricated using a CNT ink printed on normal papers.

**Figure captions**

Fig. 1 SEM image of the starting DWCNT sample.

Fig. 2 TEM image of the oxidized DWNT sample.

Fig. 3 Raman spectrum of the oxidized DWNT sample.

Fig. 4 CNT inks.

Fig. 5  The real and imaginary part of the CNT disk A deposited on paper as a function of frequency and applied voltages.

Fig. 6 The real and imaginary part of the CNT disk E deposited on paper as a function of frequency and applied voltages.

Fig. 7 The radio on paper concept; CNT inks of various concentrations are used to implement various functions which are usually done by R, L, C components and copper.

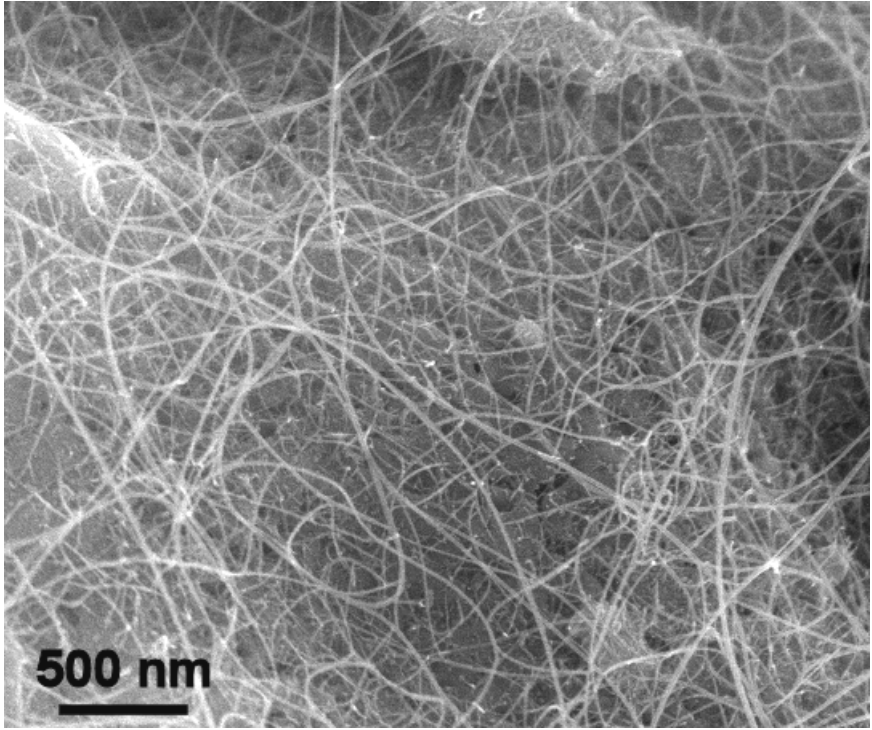

**Fig. 1** SEM image of the starting DWCNT sample.

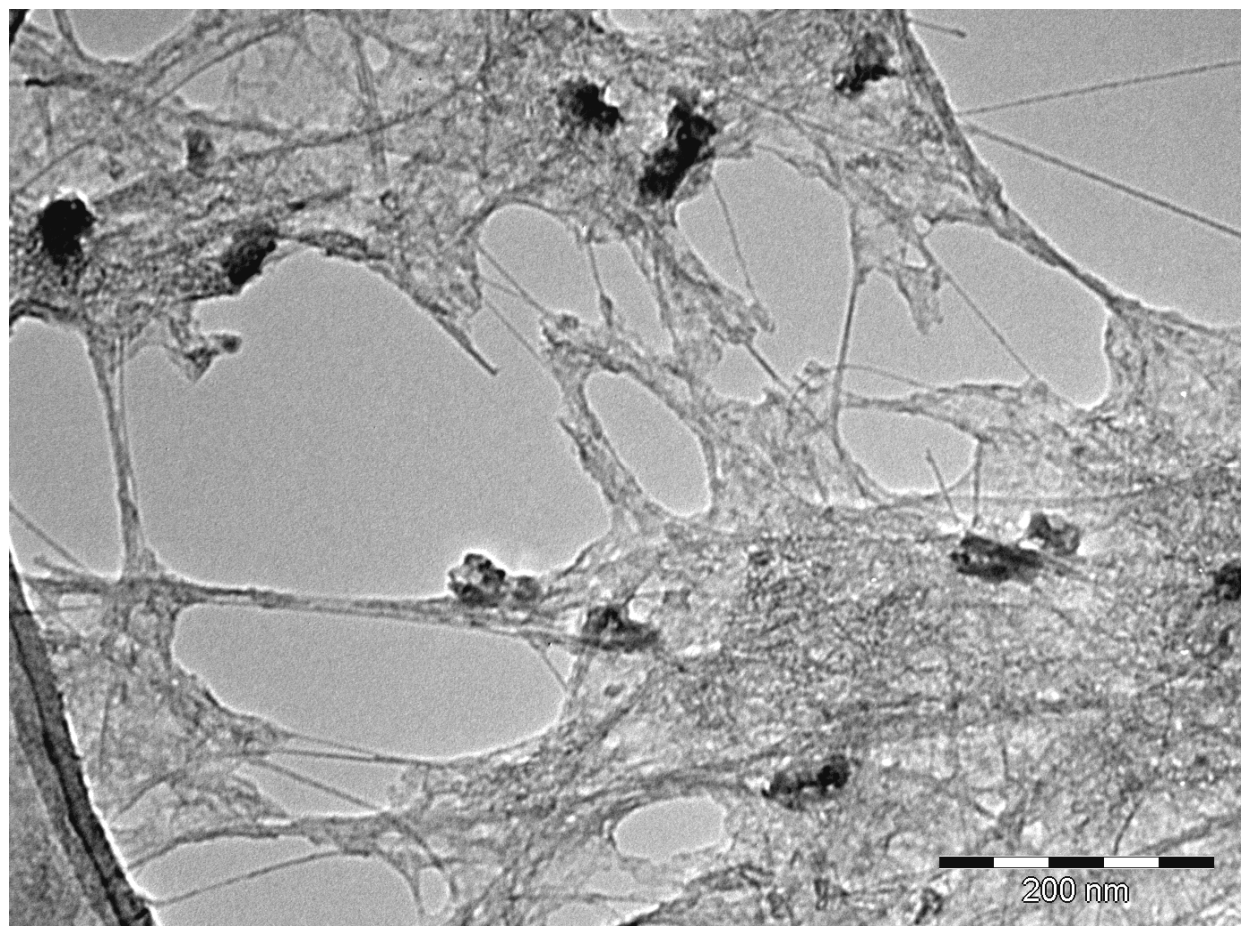

**Fig. 2** TEM image of the oxidized DWNT sample.

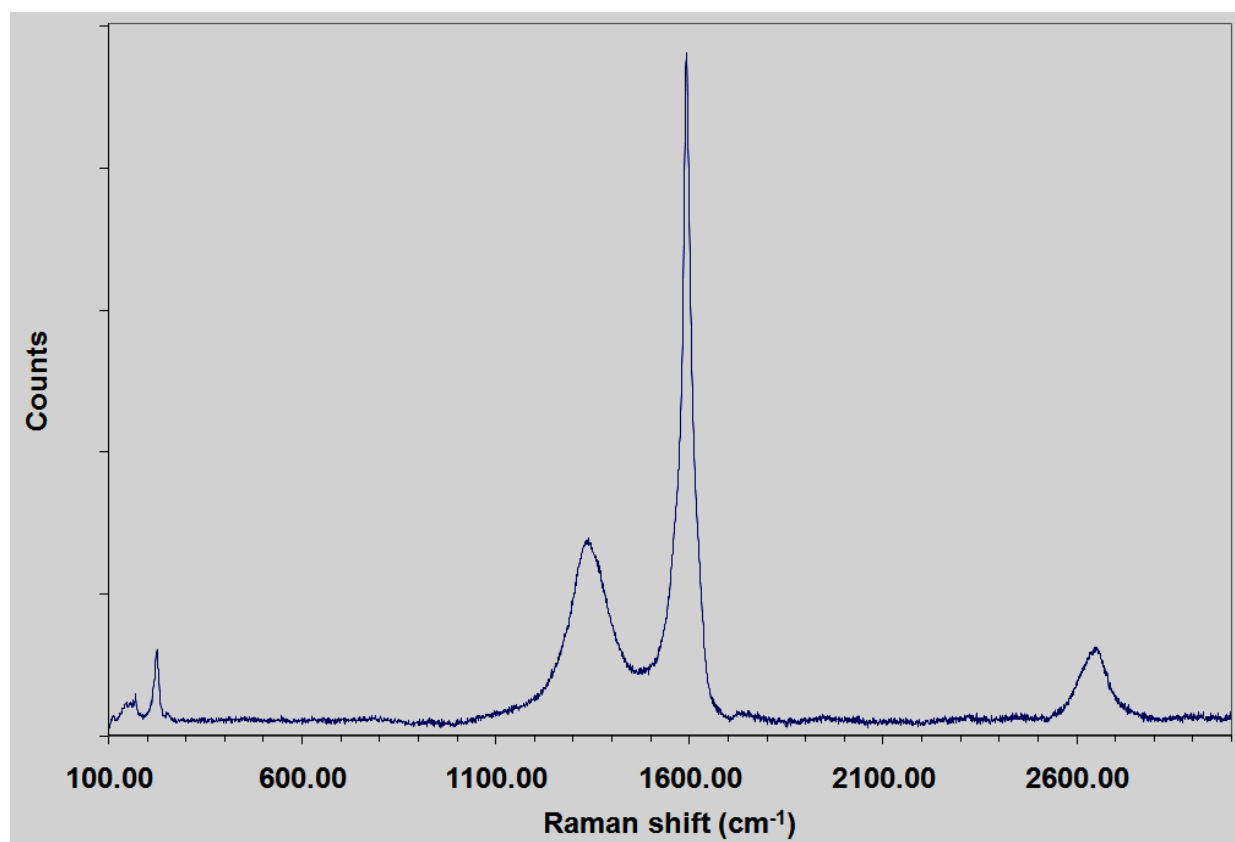

**Fig. 3** Raman spectrum of the oxidized DWNT sample.

12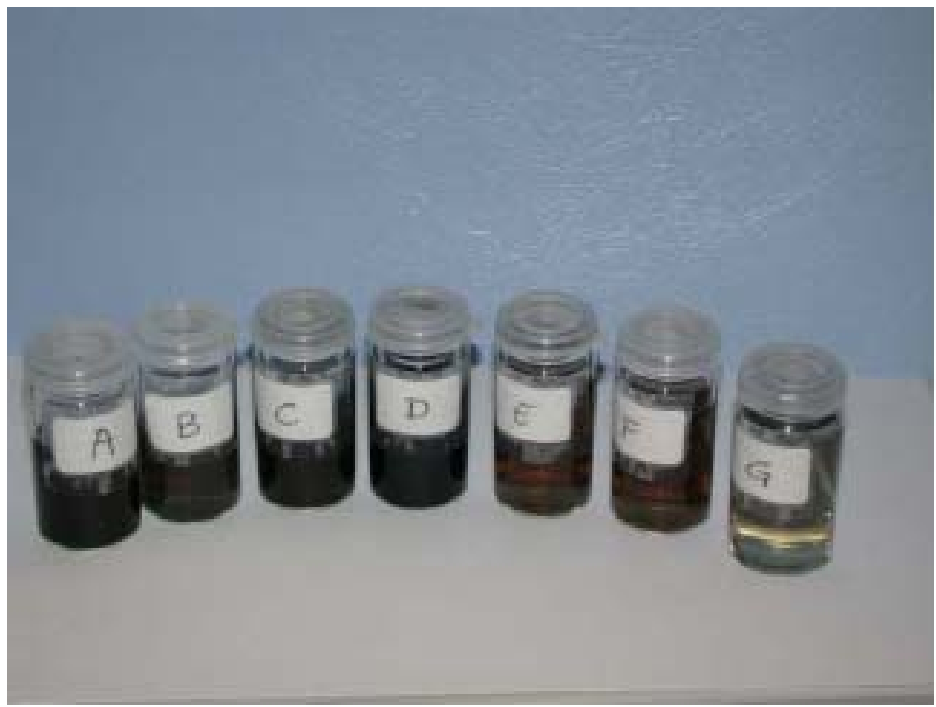

**Fig. 4** CNT inks.

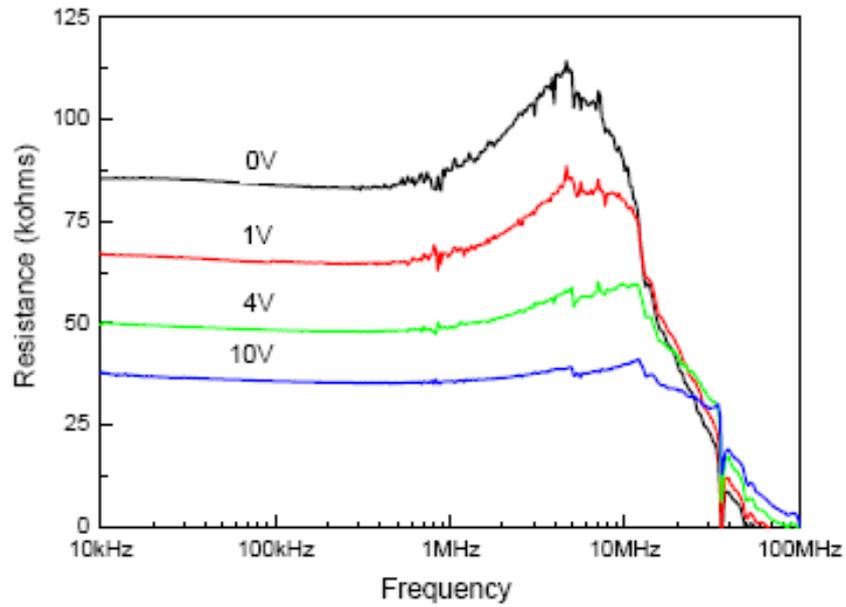

**a**

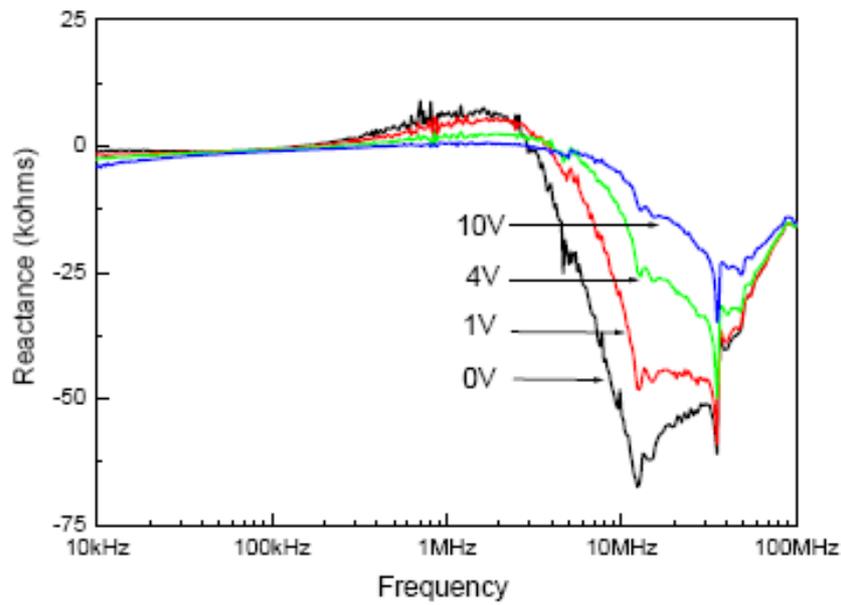

**(b)**

**Fig. 5** The real and imaginary part of the CNT disk A deposited on paper as a function of frequency and applied voltages.

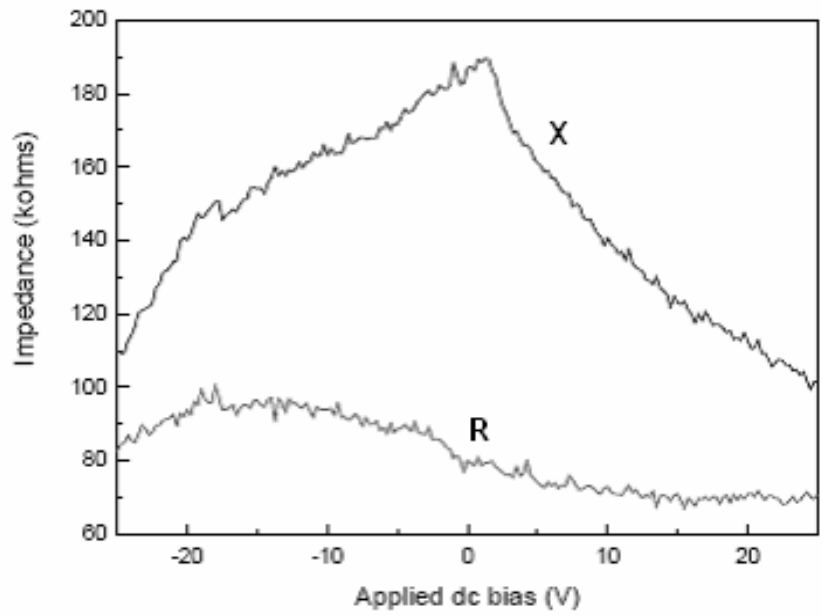

**Fig. 6** The real and imaginary part of the CNT disk E deposited on paper as a function of frequency and applied voltages.

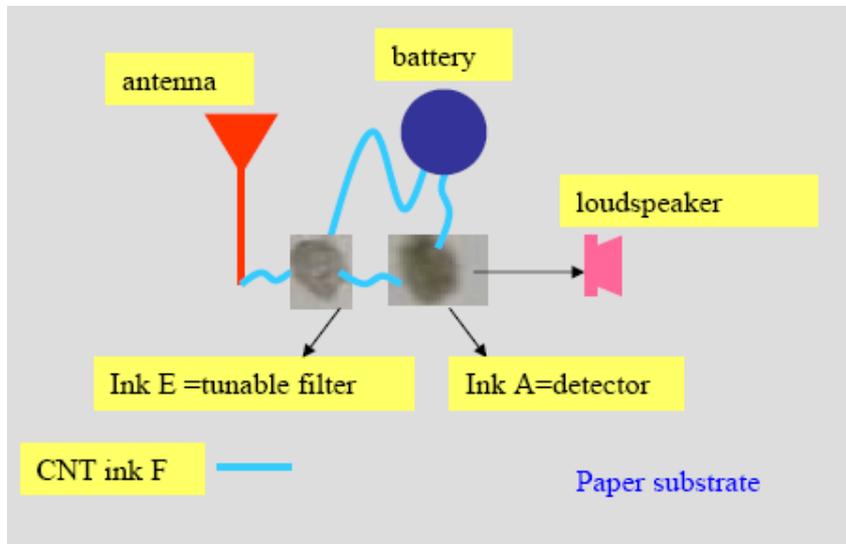

Fig. 7 The radio on paper concept; CNT inks of various concentrations are used to implement various functions which are usually done by R, L, C components and copper.